\documentclass[twocolumn,showpacs,prl,longbibliography]{revtex4-2}%
\usepackage[text={18cm,24cm}]{geometry}

\usepackage[normalem]{ulem}
\usepackage{xcolor}
\usepackage{titlesec}

\usepackage{amsfonts}
\usepackage{amsmath, bm}
\usepackage{amssymb}
\usepackage{subfigure}
\usepackage{graphicx}
\usepackage{txfonts}
\usepackage{epstopdf}
\usepackage{makecell}
\usepackage[mathlines]{lineno}
\usepackage[colorlinks,linkcolor=blue,citecolor=blue,urlcolor=blue]{hyperref}
\usepackage{amsmath}%
\providecommand{\U}[1]{\protect\rule{.1in}{.1in}}

\begin{document}

\title{Antiferromagnetic pseudospintronics without spin splitting}
\author{Shu-Hui Zhang$^{1}$}
\email{shuhuizhang@nankai.edu.cn}
\author{Shu-Qi Liu$^{2}$}
\author{Dan-Yang Han$^{3}$}
\author{Zhan Kong$^{4}$}
\author{Ding-Fu Shao$^{3}$}
\email{dfshao@issp.ac.cn}
\author{Wen Yang$^{4}$}
\author{Kai Chang$^{5,6}$}
\email{kchang@zju.edu.cn}

\affiliation{$^{1}$School of Physics, Nankai University, Tianjin 300071, China}
\affiliation{$^{2}$College of Mathematics and Physics, Beijing University of Chemical Technology, Beijing 100029, China}
\affiliation{$^{3}$Key Laboratory of Materials Physics, Institute of Solid State Physics, HFIPS, Chinese Academy of Sciences,
Hefei 230031, China}
\affiliation{$^{4}$Beijing Computational Science Research Center, Beijing 100193, China}
\affiliation{$^{5}$Center for Quantum Matter, School of Physics, Zhejiang University, Hangzhou 310027, China}
\affiliation{$^{6}$Institute for Advanced Study in Physics, Zhejiang University, Hangzhou 310027, China}

\begin{abstract}
Antiferromagnets (AFMs) are promising for high‑density spintronics due to their zero net magnetization, yet conventional AFM spintronics relies on spin splitting—a requirement that excludes many collinear AFMs with compensated spin sublattices. Here we exploit the sublattice degree of freedom in a honeycomb AFM with zero spin splitting. We uncover a coupling between spin and sublattice: the out‑of‑plane pseudospin polarization is spin‑dependent, a mechanism we term partial pseudospin–spin coupling. This allows switching of the pseudospin polarization by reversing the Néel vector. Introducing an impurity into a specific sublattice induces Friedel oscillations with a sublattice‑resolved amplitude ratio dictated solely by the pseudospin polarization, which is directly measurable by spin‑polarized scanning tunneling microscopy. Furthermore, we demonstrate Néel‑vector‑controlled transmission and a large nonvolatile tunneling magnetoresistance in an all‑in‑one AFM junction, with pronounced resonant enhancement in gate‑tunable two‑dimensional devices. Our work establishes a new paradigm—AFM pseudospintronics—that utilizes the sublattice pseudospin in zero‑spin‑splitting AFMs, extending spintronics beyond the conventional spin‑splitting paradigm.
\end{abstract}
\maketitle

\textit{Introduction}.---Spintronics utilizes the spin degree of freedom to
build quantum devices, for which the antiferromagnet (AFMs) are promising platforms due to zero net magnetization, i.e., AFM
spintronics\cite{nnano.2016.18,RevModPhys.90.015005,s41567-018-0063-6,s41567-018-0051-x,s41563-023-01492-6}. Typically, the spin splitting for AFM spintronics is required, e.g.,
altermagnetic splitting\cite{PhysRevX.12.031042,PhysRevX.12.040501} in
momentum space. However, the collinear AFM, featured by zero spin splitting due to
the parity-time ($\mathcal{PT}$) symmetry, is not viable for spintronics. One
direction is to persist with the spin degree of freedom of the collinear AFM,
which necessitates inducing spin splitting, as exemplified by three recent
works utilizing Floquet engineering\cite{7ywb-ml2q,9346-9jpf,xzm1-l6yf}.
Beyond the spin degree of freedom itself, antiferromagnets (even collinear ones) also
possess a sublattice degree of freedom. Yet, the role of sublattice
pseudospin in AFM-based applications remains largely unexplored.

In a honeycomb lattice, the sublattice pseudospin associated with the
inequivalent $A$ and $B$ sublattices underlies many of the remarkable
transport phenomena in graphene\cite{nmat3305}. The pseudospin-momentum
locking suppresses disorder-induced backscattering\cite{Ando1998} and enables
Klein tunneling through pseudospin conservation of Dirac
fermions\cite{nphys384}. Moreover, treating the sublattice pseudospin as an
analogue of the real spin enables pseudospin-based valve effects with giant
pseudomagnetoresistance, opening the avenue toward graphene
pseudospintronics\cite{PhysRevLett.102.247204}. However, existing
graphene-based pseudospintronic proposals rely on electrostatically induced
pseudospin polarization and are therefore inherently volatile. AFMs,
intrinsically hosting sublattice degrees of freedom, thus provide a natural
route toward nonvolatile pseudospintronics, i.e., AFM pseudospintronics.

Here, using a minimal model of a honeycomb antiferromagnet (AFM), we elucidate
the role of sublattice pseudospin in AFM pseudospintronics. The essential
physics originates from the partial pseudospin-spin coupling, which gives rise
to opposite out-of-plane pseudospin polarizations for the two spin branches,
despite the zero spin splitting enforced by combined $\mathcal{PT}$ symmetry.
We show that such pseudospin polarization can be directly characterized
through sublattice-resolved Friedel oscillations (FOs) induced by impurities
on inequivalent sublattices, whose relative amplitudes are uniquely determined
by the pseudospin polarization (i.e., the intrinsic amplitude ratio) and can
therefore be detected by spin-polarized scanning tunneling microscopy. More
importantly, reversing the Néel vector (NV) switches the pseudospin
polarization, enabling tunable matching and mismatching of pseudospin
polarization across AFM junctions. Consequently, we demonstrate NV-controlled
transmission and a large nonvolatile pseudomagnetoresistance effect in an
all-in-one AFM junction. The gate-tunable two-dimensional junction further
supports resonant enhancement of the pseudomagnetoresistance, facilitating its
experimental observation. Our work establishes a general framework for
exploiting sublattice pseudospin in AFM pseudospintronics.

\begin{figure}[ptbh]
\includegraphics[width=1.0\columnwidth,clip]{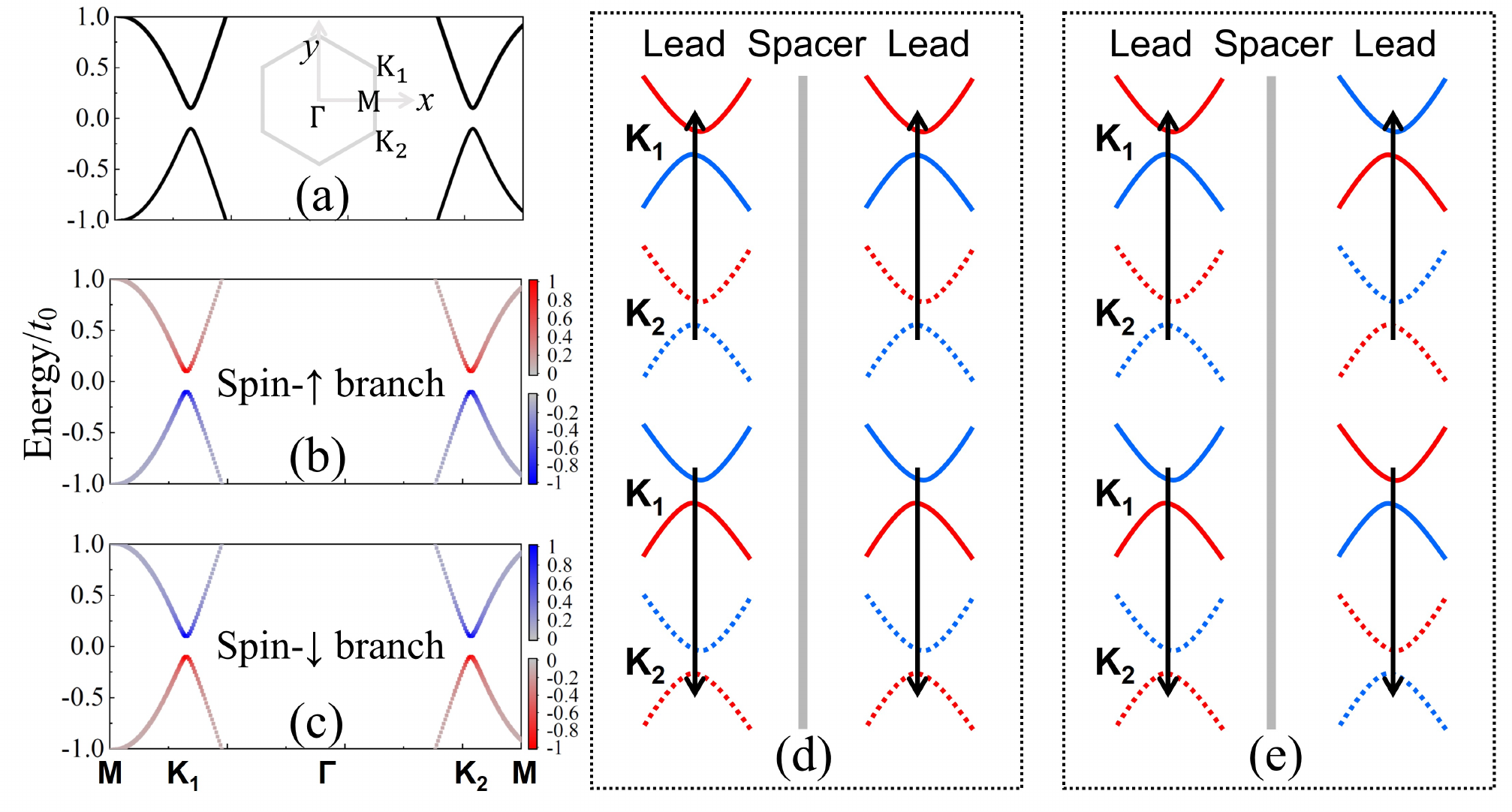}\caption{(a) The
spin-degenerate band structure of electrons in the honeycomb AFM. The inset
shows the high-symmetry points in the Brillouin zone. (b, c) Two spin branches
are associated with the pseudospin polarization $\bar{\sigma}_{\lambda s}^{z}$
(color scale), which are opposite. (d, e) The band matching of AFM-based
tunneling junctions with the junction interfaces along the zigzag directions.
The N\'{e}el vectors in the two leads are identical/opposite in (d)/(e). }%
\label{ESP}%
\end{figure}

\textit{Model}.---To consider the honeycomb AFM with the collinear
antiferromagnetic order, the tight-binding Hamiltonian can be written
as\cite{7ywb-ml2q,9346-9jpf,xzm1-l6yf}
\begin{equation}
H(\mathbf{k})=\left[
\begin{array}
[c]{cc}%
0 & f_{\mathbf{k}}\\
f_{\mathbf{k}}^{\ast} & 0
\end{array}
\right]  +\Lambda\mathbf{n\cdot\hat{\tau}}\otimes\hat{\sigma}_{z},
\label{AFMH}%
\end{equation}
by using the sublattice basis of \{$A\uparrow,B\uparrow,A\downarrow
,B\downarrow$\}. Here $f_{\mathbf{k}}=t_{0}\sum_{i=1}^{3}e^{i\mathbf{k\cdot
\delta}_{i}}$ with $\delta_{1}=a_{0}(1,0)$, $\delta_{2,3}=a_{0}(-1/2,\pm
\sqrt{3}/2)$, the exchange splitting $\Lambda$ is induced by the
antiferromagnetic order with the NV along the $\mathbf{n}$, $\mathbf{\hat
{\tau}}$ and $\mathbf{\hat{\sigma}}$ are the Pauli matrices acting on the real
spin and sublattice pseudospin, respectively. $t_{0}$ ($a_{0}$) is used as the
energy (length) later. Assuming $\mathbf{n}$ along the $z$ direction, the
electronic dispersion is%
\begin{equation}
\varepsilon_{\lambda s}(\mathbf{k})=\lambda\sqrt{\left\vert f(\mathbf{k}%
)\right\vert ^{2}+\Lambda_{s}^{2}},
\end{equation}
which is spin-degenerate, ensured by the combined $\mathcal{PT}$ symmetry as
shown by Fig. \ref{ESP}(a). Here, $\Lambda_{s}=s\Lambda$, where $s=\pm1$
labels the spin-up and spin-down branches, respectively. The wave function is
\begin{equation}
\psi_{\lambda s}(\mathbf{k})=\left[
\begin{array}
[c]{c}%
\sqrt{\frac{\varepsilon_{\lambda s}+\Lambda_{s}}{2\varepsilon_{\lambda s}}}\\
\lambda\sqrt{\frac{\varepsilon_{\lambda s}-\Lambda_{s}}{2\varepsilon_{\lambda
s}}}\frac{\left\vert f(\mathbf{k})\right\vert }{f_{\mathbf{k}}}%
\end{array}
\right]  , \label{TBWS}%
\end{equation}
which helps to define the sublattice pseudospin vector:
\begin{equation}
\bm{\bar{\sigma}}_{\lambda s}(\mathbf{k})\mathbf{=}\left\langle \psi_{\lambda
s}\right\vert \mathbf{\hat{\sigma}}\left\vert \psi_{\lambda s}\right\rangle
=\frac{(\operatorname{Re}[f^{\ast}(\mathbf{k})],\operatorname{Im}[f^{\ast
}(\mathbf{k})],\Lambda_{s})}{\varepsilon_{\lambda s}}.
\end{equation}
Although $\varepsilon_{\lambda s}\ $and $\psi_{\lambda s}$ are necessarily
associated with the spin index $s$, the explicit dependence on $s$ appears
only in the sublattice amplitudes of $\psi_{\lambda s}$. Thus, only
$\bar{\sigma}_{\lambda s}^{z}$ is spin-dependent, i.e., partial
pseudospin-spin coupling. Two spin branches have the opposite pseudospin
polarization $\bar{\sigma}_{\lambda s}^{z}$ as shown by Fig. \ref{ESP}(b, c),
which can be reversed by the NV of AFM. Notably, spin and pseudospin are
associated with $\mathcal{T}$ symmetry and $\mathcal{P}$ symmetry,
respectively. Therefore, reversing the NV of the AFM is equivalent to applying
a $\mathcal{T}$ operation, which consequently reverses the pseudospin
polarization for each spin branch since no $\mathcal{P}$ symmetry is applied.
Thus, if $\Lambda=0$, this type of pseudospin-spin coupling vanishes.
Conversely, $\Lambda\neq0$ highlights the potential of honeycomb AFMs
with spin-degenerate band structures, without requiring spin splitting.

\begin{figure}[ptbh]
\includegraphics[width=1.0\columnwidth,clip]{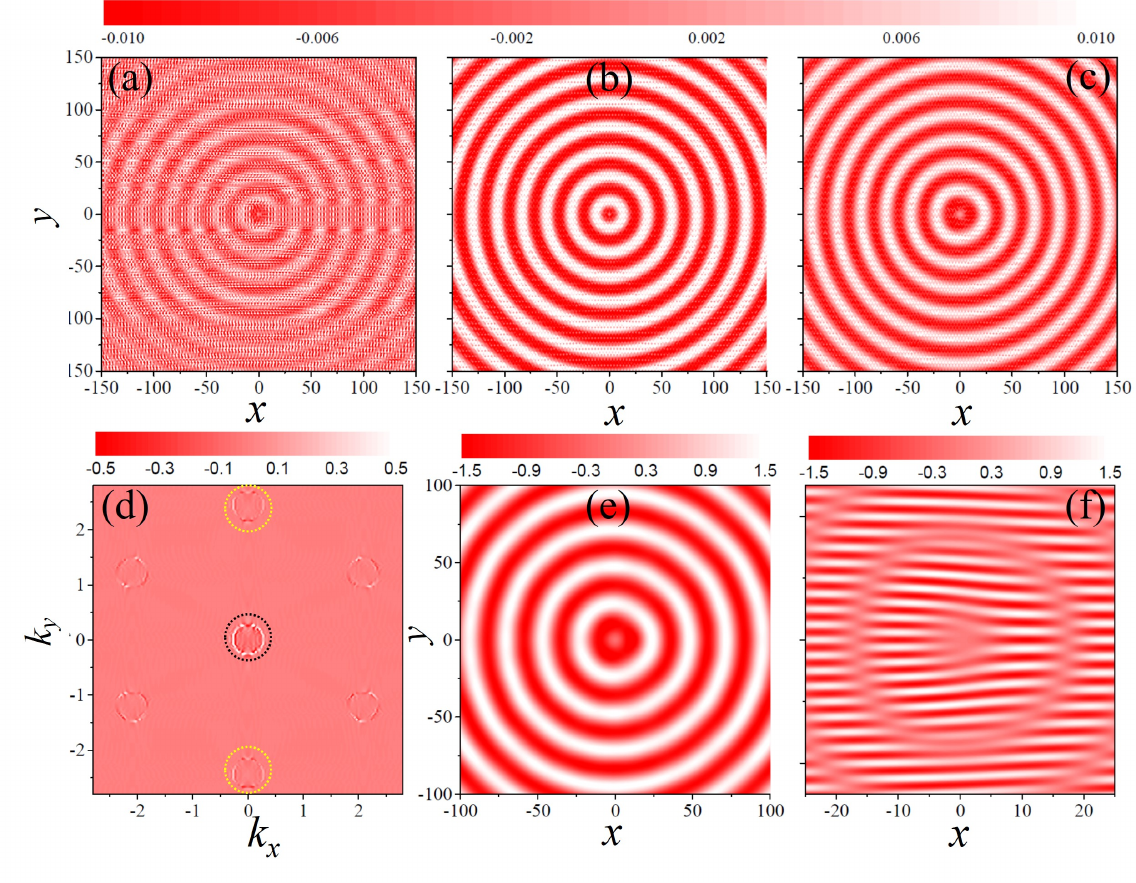}\caption{(a)
FOs induced by the impurity at the sublattice $A$ with $\mathbf{r=0}$. (b) and
(c) for the sublattice-resolved FOs at $A$ and $B$, respectively. (d) Fourier
transform of the sublattice-resolved FOs at $B$. (e) FOs contributed by the
intravalley scattering $r\delta\rho_{B}^{\text{intra}}\times10^{2}$\ through
the inverse Fourier transform by considering the momentum region limited by
the black circle in (d). (f) FOs contributed by the intervalley scattering
$r\delta\rho_{B}^{\text{inter}}\times10^{3}$ through the inverse Fourier
transform by considering the momentum region limited by yellow circles\ in
(d). Here, $\varepsilon=0.2t_{0}$, $\Delta_{0}=0.03t_{0}$, and the impurity
strength $V_{\text{imp}}=t_{0}$. }%
\label{FO_contour}%
\end{figure}

\textit{Pseudospin polarization characterization}.---Fully characterizing the
pseudospin polarization vector $\bm{\bar{\sigma}}_{\lambda s}$ is generally
challenging. If STM were pseudospin-resolved, this task could, in principle,
be addressed following the recent proposal in Ref.~\cite{nzc1-wg6m}. Although
current STM techniques are not pseudospin-resolved, they are nevertheless
sufficient to characterize the crucial quantity $\bar{\sigma}_{\lambda s}^{z}$
relevant to the present study. We first demonstrate this analytically and then
further confirm it through numerical calculations. For the honeycomb AFM, the
low-energy Hamiltonian is\cite{PS-SM}
\begin{equation}
H_{\eta s}(\mathbf{q})=v_{F}\left(  \eta\sigma_{x}q_{x}+\sigma_{y}%
q_{y}\right)  +\Lambda_{s}\sigma_{z}. \label{CM}%
\end{equation}
By intentionally placing an impurity on a specific sublattice, one can induce
the well-known FOs\cite{PhysRevLett.100.076601}. Subsequent STM measurements
may reveal the geometric Berry phase or the topological winding number through
the filtered Fourier analysis of the FOs arising from the intervalley
scattering\cite{s41586-019-1613-5,PhysRevLett.125.116804,PhysRevB.103.L161407,PhysRevB.104.035402}%
. However, an analogous filtered Fourier analysis for the FOs originating from
intravalley scattering is still lacking, which is precisely the focus of the
present study. For Eq. (\ref{CM}), the sublattice-resolved FOs contributed by
intravalley scattering take the form\cite{PS-SM}:
\begin{equation}
\delta\rho_{A/B}^{\text{intra}}=\sum_{s=\pm}\delta\rho_{s,A/B}^{\text{intra}},
\end{equation}
where
\begin{subequations}
\begin{align}
\delta\rho_{s,A}^{\text{intra}}  &  \approx-\frac{C_{0}}{r}\epsilon
_{s,+}^{3/2}\epsilon_{s,-}^{-1/2}\operatorname{Im}\left[  it_{s}%
(\varepsilon)e^{i2u_{s}}\right]  ,\\
\delta\rho_{s,B}^{\text{intra}}  &  \approx\frac{C_{0}}{r}\epsilon_{s,+}%
^{1/2}\epsilon_{s,-}^{1/2}\operatorname{Im}\left[  t_{s}(\varepsilon
)ie^{i2u_{s}}\right]  .
\end{align}
The subscript $s$ enters the above expression through $\varepsilon_{s,\pm
}=\varepsilon\pm\Lambda_{s}$ with $\varepsilon$ the Fermi level. To perform the
spin-resolved STM measurement, the spin-resolved amplitude ratio will be
\end{subequations}
\begin{equation}
R_{s}=\frac{\delta\rho_{s,A}^{\text{intra}}}{\delta\rho_{s,B}^{\text{intra}}%
}=\frac{\bar{\sigma}_{s}^{z}+1}{\bar{\sigma}_{s}^{z}-1}. \label{ARS}%
\end{equation}
Here, $R_{s}$ is independent of the specific delta-type impurity $V_{\text{imp}}\delta(\mathbf{r})$ considered since the $T$-matrix $t_{s}(\varepsilon)$ occurs
identically in $\delta\rho_{s,A/B}^{\text{intra}}$. Remarkably, $R_{s}$
directly gives the pseudospin polarization $\bar{\sigma}_{s}^{z}=$
$\Lambda_{s}/\varepsilon$. Here, $\lambda$ is omitted for brevity.
Alternatively, the pseudospin polarization characterization can also be
realized through the spin- and sublattice-resolved impurity plus the
conventional sublattice-resolved STM measurement.

\begin{figure}[ptbh]
\includegraphics[width=1.0\columnwidth,clip]{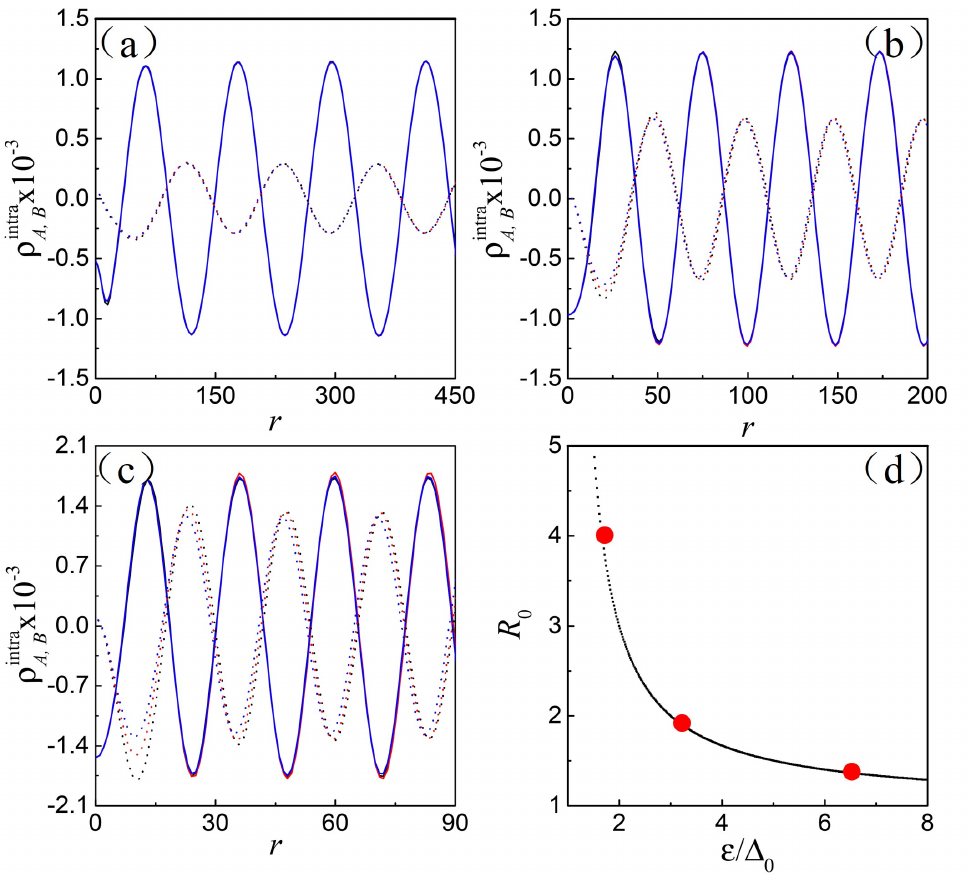}\caption{$\delta
\rho_{A,B}^{\text{intra}}$ and the amplitude ratio $R=\delta\rho
_{A}^{\text{intra}}/\delta\rho_{B}^{\text{intra}}$ from the inverse Fourier
transform. (a) $\varepsilon=0.05t_{0}$. (b) $\varepsilon=0.1t_{0}$. (c)
$\varepsilon=0.2t_{0}$.\ (d) $R$ as the function of $\varepsilon/\Delta_{0}%
$.\ In (a, b, c), three directions $\theta_{\mathbf{r}}=0$ (black), $\pi/6$
(red), $\pi/3$ (blue) are considered, and the solid (dotted)\ lines are for
$\delta\rho_{A}^{\text{intra}}$ ($\delta\rho_{B}^{\text{intra}}$). Here,
$\Delta_{0}=0.03t_{0}$. }%
\label{rat}%
\end{figure}
However, the STM measurement is not straightforward. Based on the
tight-binding Hamiltonian, Eq. (\ref{AFMH}), we numerically demonstrate the
extraction procedure. The spin-polarized STM simulation first yields Fig.
\ref{FO_contour}(a), which can be separately resolved on the $A$ and $B$
sublattices, as shown in Figs. \ref{FO_contour}(b) and \ref{FO_contour}(c),
respectively. Here, only the $s=+$ branch is considered, and the spin index is
omitted hereafter for brevity. Performing the Fourier transform of the
sublattice-resolved FOs, e.g., for Fig. \ref{FO_contour}(c), gives Fig.
\ref{FO_contour}(d), where the intravalley and intervalley scattering
processes are clearly distinguished. By subsequently carrying out the inverse
Fourier transform with momentum-space filtering, namely selecting the regions
enclosed by the red circle (yellow circles) corresponding to the intravalley
(intervalley) scattering, one obtains Figs. \ref{FO_contour}(e) and
\ref{FO_contour}(f), respectively. Figure \ref{FO_contour}(f) exhibits the
well-known pair of wavefront
dislocations\cite{PhysRevB.103.L161407,PhysRevB.104.035402}, originating from
the topological winding number of electrons in the honeycomb AFM. In contrast,
Fig. \ref{FO_contour}(e) represents the FOs arising solely from the
intravalley scattering process, from which the ratio $R$ can be extracted directly.

Figure \ref{rat} presents the extracted $\delta\rho_{A,B}^{\text{intra}}$ and
the corresponding ratio $R$. Three representative directions are considered,
namely $\theta_{\mathbf{r}}=0$, $\pi/6$, and $\pi/3$. As shown in Fig.
\ref{rat}(a), the curves for $\theta_{\mathbf{r}}=0$ and $\pi/3$ coincide,
reflecting the $C_{6}$ rotational symmetry of the sublattices in the honeycomb
AFM. At low doping, $\varepsilon=0.05t_{0}$ [Fig. \ref{rat}(a)], the curve for
$\theta_{\mathbf{r}}=\pi/6$ also nearly overlaps with the other two due to the
negligible trigonal warping effect\cite{PhysRevB.97.035420}. As the doping
level $\varepsilon$ increases, the trigonal warping gradually becomes
pronounced, causing the $\theta_{\mathbf{r}}=\pi/6$ curve to separate from the
other two. Even at $\varepsilon=0.2t_{0}$ [cf. Fig. \ref{rat}(c)], the
extracted ratio $R$ remains in excellent agreement with the analytical
expression in Eq. (\ref{ARS}), with deviations below $1\%$, as demonstrated in
Fig. \ref{rat}(d). The remarkable validity of Eq. (\ref{ARS}), derived from
the low-energy Hamiltonian, originates from the tight-binding wave function
[cf. Eq. (\ref{TBWS})], whose component ratio is independent of the trigonal
warping effect.

\begin{figure}[ptbh]
\includegraphics[width=1.0\columnwidth,clip]{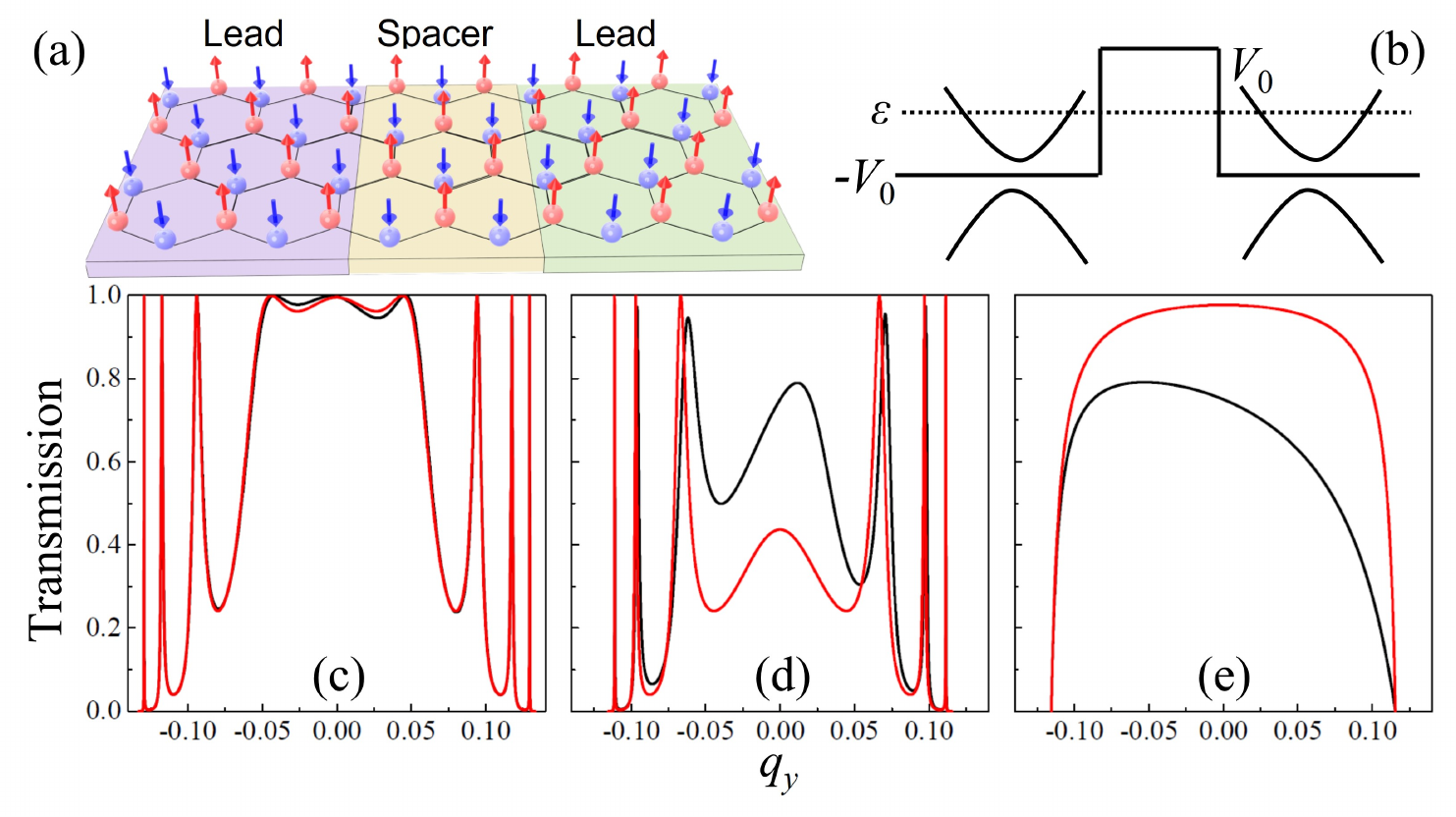}\caption{An
all-in-one junction based on the honeycomb AFM (a) and the corresponding
potential distribution (b). (c, d, e) The change of transmission induced by the
inversion of the Neel vector. (c) $\Lambda_{3}=\pm0.05V_{0}$ for the red solid
line and the black dotted line, respectively, and $w=100a_{0}$. (d)
$\Lambda_{3}=\pm0.5V_{0}$ for the red solid line and the black dotted line,
respectively, and $w=100a_{0}$. (e) $\Lambda_{3}=\pm0.5V_{0}$ for the red
solid line and the black dotted line, respectively, and $w=a_{0}$. Here, the
spin-up branch of $\mathbf{K}_{1}$ valley is considered, $\varepsilon=0$,
$V_{0}=0.2t_{0}$, and $\Lambda_{1}=\Lambda_{2}=$ $0.5V_{0}$. }%
\label{tran}%
\end{figure}

The pseudospin polarization characterization via spin-polarized STM is
generally applicable to both spin branches. Since $\bar{\sigma}_{s}%
^{z}=\Lambda_{s}/\varepsilon$ reverses with the NV of the AFM, the proposed
pseudospin-polarization characterization simultaneously enables local
detection of the NV, which is of central importance in AFM
spintronics\cite{RevModPhys.90.015005}. In realistic AFMs, different magnetic
domains typically exist on the micrometer scale. Existing global measurement
techniques\cite{science.aab1031,s41567-018-0062-7,PhysRevLett.124.067203,PhysRevLett.127.277201,PhysRevLett.127.277202,PhysRevLett.131.056401,s41586-021-03679-w,s41467-021-21872-3,adn0479,PhysRevLett.133.096803}
are capable of determining the NV in a single-domain sample, but are less
suitable for characterizing AFMs containing multiple domains. Therefore, our
proposed local detection scheme for the NV is particularly advantageous for
probing the NV textures across multiple AFM domain
walls\cite{s41586-024-08234-x}.

\textit{Nonvolatile pseudomagnetoresistance}.---Tunneling magnetoresistance is
a fundamental function required by magnetic devices. Exploiting the NV
controlled pseudospin polarization, junctions based on the honeycomb AFM
should exhibit the magnetoresistance effect, i.e.,
pseudomagnetoresistance\cite{PhysRevLett.102.247204}. In Fig. \ref{tran}(a),
the conventional lead-spacer-lead geometry is considered, which forms an
all-in-one junction based on the honeycomb AFM. Figure \ref{tran}(b) schematically shows the potential profile, where $-V_{0}$ ($V_{0}$) is applied to the lead (spacer) regions via the back and top gates in the all-in-one AFM junction. The electrons of honeycomb AFM feature multiple internal degrees of freedom,
such as valley, spin, and sublattice pseudospin. To highlight the sublattice
pseudospin, we assume that the junction interfaces are along the zigzag
directions of honeycomb AFM, preventing the intervalley
coupling\cite{PhysRevB.97.035420} and the barriers do not induce spin
mixing. Thus, the spin and valley are conserved in tunneling processes. The
conservation of sublattice pseudospin may generate the Klein tunneling or
anti-Klein tunneling\cite{nphys384}, but the sublattice pseudospin is not
necessarily conserved. Switching the NV in the right lead reverses the
spin-dependent gap and then $\bar{\sigma}_{s}^{z}$, leading to the different
band matching in Fig. \ref{ESP}(d) and (e).

\begin{figure}[ptbh]
\includegraphics[width=1.0\columnwidth,clip]{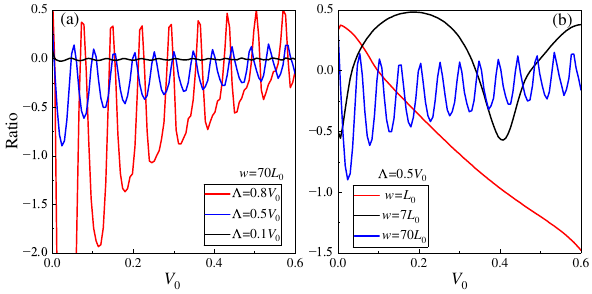}\caption{The
pseudomagnetoresistance ratio as a function of junction potential $V_{0}$.
Different gap values $\Lambda$ and junction widths $w$ are considered. Here,
$L_{0}=1.5a_{0}$. }%
\label{ratio}%
\end{figure}

To consider the change of transmission induced by the reversion of the NV,
Fig. \ref{tran} is plotted for the spin-up branch of $\mathbf{K}_{1}$ valley,
i.e., the first row in Fig. \ref{ESP}(b) and (c). Here, for analytical
derivations, the low-energy model of Eq. (\ref{CM}) is used, which favors the
explicit expression for the red lines but not for the black lines [cf.
SM\cite{PS-SM}]. In Fig. \ref{tran}(c), for the right lead we adopt the gap
value $\Lambda_{3}=\pm0.05V_{0}$ and the barrier width $w=100a_{0}$. Note
that $\Lambda$ is associated with the subscript $1$, $2$, $3$ for the junction
regions from left to right.$\ $The Fermi level is $\varepsilon=0$, but the
doping is nonzero, e.g., the doping is $\varepsilon\pm V_{0}$ in the lead and
spacer regions, respectively. Because the Fermi level is away from the band
bottom, the gap effect or the pseudospin polarization effect is trivial for
the transmission [cf. nearly overlapping lines in Fig. \ref{tran}(c)]. In
addition, the perfect transmission peaks up to $1$ occur due to the resonant
matching of the longitudinal electronic wavelength with the barrier width
$w$\cite{PhysRevLett.131.246301}. By using $\Lambda_{3}=\pm0.5V_{0}$, the
large transmission difference can be seen in Fig. \ref{tran}(d). Particularly,
the higher transmission takes place for the opposite NVs in the two leads [cf. the
black line in Fig. \ref{tran}(d)], in contrast to the cases of AFM junctions
by using the spin matching\cite{PhysRevLett.130.216702}. The reason is the
competition between the pseudospin matching and the resonant tunneling. To
clearly show this point, the short junction with an impractical width
$w=a_{0}$ is considered in Fig. \ref{tran}(e), which suppresses the resonant
tunneling and highlights the role of pseudospin matching, see the red line
in Fig. \ref{tran}(e) for the higher electronic transmission across the
junction with the parallel NVs in the two leads.

Using the transmission, the conductance between two leads can be calculated,
e.g., $\sigma_{P}$ ($\sigma_{AP}$) for two leads with the parallel
(antiparallel)\ NVs. Then, the pseudomagnetoresistance ratio can be defined
as\cite{PhysRevLett.102.247204}:
\begin{equation}
\text{PMR}=\frac{\sigma_{P}-\sigma_{AP}}{\sigma_{P}}.
\end{equation}
For the exact calculations, the tight-binding Hamiltonian of Eq. (\ref{AFMH})
is used [cf. SM\cite{PS-SM}]. The pseudomagnetoresistance ratio is shown by
Fig. \ref{ratio} as a function of junction potential $V_{0}$. Different
gap values $\Lambda=\Lambda_{1}=\Lambda_{2}=\left\vert \Lambda_{3}\right\vert
$ and junction widths $w$ are considered. In Fig. \ref{ratio}(a), with
increasing $\Lambda$, the pseudomagnetoresistance ratio becomes larger, which
is consistent with the transmission discussions. Following the transmission,
the pseudomagnetoresistance ratio is also oscillatory. If the junction width
is decreased, the pseudomagnetoresistance ratio loses its oscillatory behavior as shown by Fig. \ref{ratio}(b). Obviously, the
pseudomagnetoresistance ratio is large and can be up to $100\%$, and can be
further tuned by the resonant tunneling. Compared with conventional
pseudomagnetoresistance in graphene\cite{PhysRevLett.102.247204}, the
pseudomagnetoresistance in AFM is nonvolatile, making pseudomagnetoresistance a novel mechanism to realize the AFM pseudospintronics.

\textit{Theoretical generalization and experimental feasibility}.---The partial pseudospin–spin coupling term of the form $\sim \hat{\tau}_{z}\otimes\hat{\sigma}_{z}$ also appears in other $\mathcal{PT}$-symmetric AFMs, such as square-lattice systems with nonsymmorphic symmetry\cite{PhysRevLett.124.066401,PhysRevLett.115.126803,PhysRevB.95.115138}. In contrast, the minimal model for tetragonal CuMnAs exhibits a different type of partial pseudospin–spin coupling\cite{PhysRevLett.118.106402,PhysRevX.11.011001}, namely $\sim (\hat{\tau}_{y}\sin k_{x}-\hat{\tau}_{x}\sin k_{y})\otimes\hat{\sigma}_{z}$. Such a coupling is likewise expected to promote pseudomagnetoresistance through the reversal of the in-plane spin polarization via Néel-vector (NV) manipulation. More generally, pseudospin provides a simplified description of multiorbital wave functions. In particular, each sublattice in the honeycomb AFM hosts only a single orbital, making Eq. (\ref{AFMH}) a minimal model for elucidating the influence of the NV on the electronic wave function. This perspective highlights the importance of understanding AFM transport properties from the viewpoint of wave-function components. The essential point is that AFMs with different NV configurations possess distinct electronic wave functions. In addition, it would be interesting to explore pseudomagnetoresistance in three-dimensional $\mathcal{PT}$-symmetric AFMs.

The generality of our theoretical findings makes the experimental observation of the two predicted effects highly feasible. The intrinsic amplitude ratio and its characterization for pseudospin polarization rely on STM measurements. In graphene, STM has already resolved signatures associated with intervalley scattering\cite{s41586-019-1613-5,PhysRevLett.125.116804}, which naturally paves the way for detecting the pseudospin polarization encoded in intravalley scattering. To further realize spin-resolved characterization, spin-polarized STM measurements can be implemented in graphene through AFM proximity effects\cite{pnas.1219420110,PhysRevLett.112.116404,PhysRevLett.124.136403,PhysRevLett.128.106401,s41928-020-0458-0,s41467-024-48809-w,s41467-025-64555-z,s41565-021-00887-3}. These advances indicate that the proposed characterization of pseudospin polarization is experimentally accessible with existing techniques. Intrinsic two-dimensional honeycomb AFMs, such as MnPX$_3$ (X = S, Se)\cite{nanolett.9b05165,S0925838820307957}, also provide promising experimental platforms. Nonvolatile pseudomagnetoresistance represents a macroscopic transport phenomenon and should therefore be comparatively straightforward to observe experimentally, provided that deterministic switching of the NV can be achieved. In recent years, substantial theoretical\cite{fkyr-z5b8} and experimental\cite{s41565-024-01741-y,2vc9-t8qt,hv6j-vzwf,adn0479,s41586-024-08436-3} progress has been made toward deterministic control and reversal of the AFM NV, further strengthening the prospects for experimentally demonstrating the predicted nonvolatile pseudomagnetoresistance. 

\textit{Conclusions}.---In conclusion, for $\mathcal{PT}$-symmetric antiferromagnets (AFMs), we uncover the unique role of the sublattice degree of freedom in electronic scattering from zero‑dimensional impurities and one‑dimensional junctions. The underlying mechanism is partial pseudospin–spin coupling and its tunability via the Néel vector. Based on this physics, we predict two pseudospin‑related phenomena: an intrinsic amplitude ratio that characterizes pseudospin polarization, and a giant pseudomagnetoresistance. The former provides a local probe of the Néel vector, particularly advantageous for multidomain AFMs, while the latter demonstrates the potential of collinear AFMs with zero spin splitting for spintronic functionalities. Given rapid experimental progress in AFM systems, these predictions are accessible with existing techniques. Our work thus establishes a route toward AFM pseudospintronics, extending the field beyond the conventional spin‑splitting paradigm.

\textit{Acknowledgements}.---This work was supported by the Innovation Program
for Quantum Science and Technology (Grant No. 2024ZD0300104), the NSFC (Grants
No. 12174019). S.H.Z. is also supported by \textquotedblleft the Fundamental
Research Funds for the Central Universities (63263119).\textquotedblright.


\end{document}